\documentclass{article}
\usepackage[cp1251]{inputenc}
\usepackage{array,hhline,amssymb,amsthm}
\usepackage[dvips]{graphics}\usepackage{array,hhline,amssymb}
\newcounter{risunok}

\title{Spontaneous breaking of  symmetry of the
gravitons of the long wave spectrum\\ in the
early Universe.}
\author{A.~A.~Grib, V.~Yu.~Dorofeev\thanks{e-mail:
Andrei\_Grib@mail.ru, dor@vd8186.spb.edu, friedlab@mail.ru}\cr\cr
A.~A.~Friedmann laboratory for theoretical physics\\
Moika 48, St. Petersburg, Russia.}
\date{}
\begin{document}
\maketitle
\begin{abstract}
It is shown that nonlinear terms in equations of
gravitons on the background of curved space-time
of the expanding Universe can solve the problem
of the negative square of the effective mass
formally arising in linear approximation for
gravitons. Similar to well known spontaneous
breaking of symmetry in Goldstone model one must
take another vacuum so that nonzero vacuum
expectation value of the quantized graviton field
leads to change of spectrum for gravitons. There
appears two graviton fields, one with the
positive mass, another with the zero mass. Energy
density and the density of particles created by
gravitation of the expanding Universe are
calculated for some special cases of the scale
factor. Numerical of result are obtained for the
dust universe case.
\end{abstract}

\section*{Introduction}
There is a problem in quantum theory of gravitons
created from vacuum in the expanding Universe
with nonzero scalar curvature $R$ (inflation,
dust, etc.) concerning the long wave graviton
modes. In linearized theory of quantum gravitons
in curved space-time of the isotropic homogeneous
Universe one obtains after separation of
variables in the wave equation the equation for
the function dependent only on time. This
equation can be understood as equation in flat
stationary metric with time dependent mass. It
occurs that for long waves this effective mass
squared is negative. All this occurs due to
conformal noninvariance of the graviton theory
for nonzero $R$ leading to tachyonic  behaviour
of long waves modes.

In some papers \cite{Gri} (see and references
there) it was proposed to consider these modes as
classical excitations of the field growing in
time, so that one must quantize only modes with
momentum with the square larger than the negative
square of the effective mass. However one knows
from the quantum field theory that tachyonic
behaviour  disappears if one takes into account
nonlinear terms neglected in linearized theory.
This is typical in theories of spontaneous
breaking of symmetry due to redefinition of the
vacuum leading to its noninvariance to this or
that transformation of the Lagrangian.  In
quantum theory based on a new vacuum one gets new
masses for the redefined quantum field so that
there is no negative mass square.

In this paper the analogous program is made for
gravitons. It occurs that if one is going from
the linear theory of gravitons taking into
account the next order of nonlinearity one gets
the redefinition of  vacuum solving the problem
for long wave gravitons. In the result one gets
gravitons with zero and positive effective mass.
In the end of the paper the expressions for the
particle density and the energy density are
obtained for gravitons created in expanding
Universe with metric  which has some special
dependence of the scale factor on time.

\section{Getting the graviton equation from Einstein
equation}

Einstein equations in presence of matter have the
form
\begin{equation}\label{ghG}
R_{ik}-\frac12g_{ik}R=\kappa T_{ik}
\end{equation}
or
\begin{equation}\label{ghT}
R^i_k=\kappa(T^i_k-\frac12\delta^i_k T).
\end{equation}

Let us consider the case when matter is
homogeneous isotropic liquid filling the
Universe. Then
\begin{equation}\label{tei}
T_{ik}=(\varepsilon+p)u_iu_k-g_{ik}p,
\end{equation}
where $u_i$ is the four velocity, $p$ -- the
pressure and $\varepsilon$ -- the energy density
of the liquid.

The problem of creation of gravitons in the early
Universe was discussed in literature with
gravitons considered as quantized small term in
the metrical tensor. Due to absence of exact
quantum gravity usually one deals with linearized
analogy with quantization of other quantum
fields. First let us obtain equations for
classical small perturbations of the metrical
tensor and then do quantization. Consider the
graviton perturbations-the gravitational waves as
small term added to the background metric
                                                                                                                    (4)
So there are
\begin{equation}\label{fd}
g_{ik}=\stackrel{(\circ)}g_{ik}+h_{ik}
\end{equation}
if $h_{ik}=0$ the $\stackrel{(\circ)}g_{ik}$ is
the solution of Einstein's equation of the form
\begin{equation}\label{Ef}
\stackrel{(\circ)}R{}^i_k=\kappa(\stackrel{(\circ)}T{}^i_k-
\frac12\delta^i_k\stackrel{(\circ)}T).
\end{equation}

Let us go from the up to low indices and vice
verse by using the background metric
$\stackrel{(\circ)}g_{ik}:
h^i_k=\stackrel{(\circ)}g{}^{in}h_{nk}$ and
expand equations (\ref{ghT}) in a series in
$h_k^i$:
\begin{equation}\label{raz}
\stackrel{(\circ)}R{}^i_k+\delta
R^i_k=\kappa(\stackrel{(\circ)}T{}^i_k-
\frac12\delta^i_k\stackrel{(\circ)}T+\delta
T^i_k-\frac12\delta^i_k\delta T),
\end{equation}
from which due to (\ref{Ef}) the perturbations
$h_{ik}$ satisfy equations
\begin{equation}\label{uh}
\delta R^i_k=\kappa(\delta
T^i_k-\frac12\delta^i_k\delta T).
\end{equation}

Using the notation $(1+h)^{-1}$ for the matrix
inverse to $(1+h)$ with small $h_k^i$ (small in
the sense that all eigenvalues of the matrix
$(1+h)$ are smaller than the unit)one obtains
\begin{equation}\label{uh00}
{(1+h)^{-1}}^i_k=\delta^i_k-h^i_k+h^i_nh^n_k-\dots
\end{equation}

Write the Ricci tensor and the curvature tensor
as
$$R^i_k={(1+h)^{-1}}^i_{i'}(-h^{i'}_n\stackrel{(\circ)}R{}^n_k+
\frac12{(1+h)^{-1}}^l_{l'}(h^{l'i'}_{;k;l}+h^{l';i'}_{k;l}-
h^{i';l'}_{k;l}-h^{l';i'}_{l;k})+$$
$$+\frac14{(1+h)^{-1}}^l_{l'}{(1+h)^{-1}}^n_{n'}(h^{l'}_{n;k}h^{n';i'}_l
-(2h^{l'}_{n;l}-h^{l'}_{l;n}) \cdot$$
\begin{equation}\label{Ri}
\cdot(h^{n';i'}_k+h^{n'i'}_{;k}-h^{i';n'}_k
)-2h^{l'}_{k;n}(h^{n'i'}_{;l}-h^{i';n'}_l)))+\stackrel{(\circ)}R{}^i_k.
\end{equation}
$$R_{ik}=\stackrel{(\circ)}R_{ik}+
\frac12{(1+h)^{-1}}^l_{l'}(h^{l'}_{i;k;l}+h^{l'}_{k;i;l}-
h^{;l'}_{ik;l}-h^{l'}_{l;k;i})+$$
$$+\frac14{(1+h)^{-1}}^l_{l'}{(1+h)^{-1}}^n_{n'}(h^{l'}_{n;k}h^{n'}_{l;i}
-(2h^{l'}_{n;l}-h^{l'}_{l;n}) \cdot$$
\begin{equation}\label{Rik}
\cdot(h^{n'}_{k;i}+h^{n'}_{i;k}-h^{;n'}_{ik}
)-2h^{l'}_{k;n}(h^{n'}_{i;l}-h^{;n'}_{il}))).
\end{equation}
$$R=R^i_i=\stackrel{(\circ)}R+{(1+h)^{-1}}^i_{i'}
(-h^{i'}_n\stackrel{(\circ)}R{}^n_i+
{(1+h)^{-1}}^l_{l'}(h^{l';i'}_{i;l}-h^{i';l'}_{i;l})+$$
$$+\frac14{(1+h)^{-1}}^l_{l'}{(1+h)^{-1}}^n_{n'}(3h^{l'}_{n;i}h^{n';i'}_l
-2h^{l'}_{i;n}h^{n'i'}_{;l}-$$
\begin{equation}\label{R}
-h^l_{l';n}h^{i';n'}_k
+4h^{l'}_{n;l}(h^{i';n'}_i-h^{n';i'}_i )).
\end{equation}

Here ";" means the covariant derivative in
background metric $\stackrel{(\circ)}g_{ik}$.
Considering in (\ref{Rik}) only first degree in
$h_{ik}$ one obtains the linearized equations for
$h_{ik}$ in (\ref{ghG}) as
\begin{equation}\label{lho}
h^{;n}_{ik;n}+h_{;i;k}-h^{n}_{i;k;n}-h^{n}_{k;i;n}-
\stackrel{(\circ)}g_{ik}(h^{;n}_{;n}-h^{m;n}_{n;m})+
h_{ik}\stackrel{(\circ)}R=-2\kappa\delta\!\stackrel{(1)}T_{ik}.
\end{equation}

Let us consider the background as homogeneous
isotropic nonstastionary space-time
\begin{equation}\label{met}
ds^2=dt^2-a^2(t)\vec{dl}^2,
\end{equation}
where $\eta$ is the conformal time. Here the
Latin indices take the values $0,1,2,3$ and the
Greek -- $1,2,3$. Then write for the scalar
curvature and the Ricci tensor
$$\delta
R={(1+h)^{-1}}^k_{k'}({(1+h)^{-1}}^l_{l'}(h^{l'\tilde{;}k'}_{k\quad\tilde{;}l}-
h^{k'\tilde{;}l'}_{k\quad\tilde{;}l})-\frac1{a^2}h^{k'}_{k,0,0}-
\frac{3a'}{a^3}h^{k'}_{k,0}+$$
$$+\frac{2\epsilon}{a^2}h^{k'}_k+\frac14{(1+h)^{-1}}^l_{l'}
{(1+h)^{-1}}^n_{n'}(4h^{l'}_{n\tilde{;}l}(h^{k'\tilde{;}n'}_k-h^{n'\tilde{;}k'}_k)
+3h^{l'}_{k,n}h^{n'\tilde{;}k'}_l-$$
$$-2h^{l'}_{k\tilde{;}n}h^{n'k'}_{\tilde{;}l}
-h^{l'}_{l\tilde{;}n}h^{k'\tilde{;}n'}_k)+\frac1{4a^2}{(1+h)^{-1}}^l_{l'}
(3h^{l'}_{k,0}h^{k'}_{l,0}-h^{l'}_{l,0}h^{k'}_{k,0})).$$
$$\delta R^0_0=-\frac1{2a^2}{(1+h)^{-1}}^l_{l'}(h^{l'}_{l,0,0}+
\frac{a'}ah^{l'}_{l,0}-
\frac12{(1+h)^{-1}}^n_{n'}h^{l'}_{l,0}h^{n'}_{l,0}).$$
$$\delta R^0_\alpha=\frac1{2a^2}{(1+h)^{-1}}^l_{l'}
(h^{l'}_{\alpha\tilde{;}l,0}-h^{l'}_{l\tilde{;}\alpha,0})
+\frac1{4a^2}{(1+h)^{-1}}^n_{n'}\cdot$$
$$\cdot(h^{l'}_{n\tilde{;}\alpha}h^{n'}_{l,0}-h^{n'}_{\alpha,0}(2h^{l'}_{n\tilde{;}l}
-h^{l'}_{l\tilde{;}n})).$$
\begin{equation}\label{rue}
\delta R^\alpha_\beta={(1+h)^{-1}}^\alpha_{i'}
(\frac12{(1+h)^{-1}}^l_{l'}
((h^{l'i'}_{\tilde{;}\beta\tilde{;}l}+h^{l'\tilde{;}i'}_{\beta\quad\tilde{;}l}
-h^{i'\tilde{;}l'}_{\beta\quad\tilde{;}l}-h^{l'\tilde{;}i'}_{l\tilde{;}\beta})+$$
$$+\frac14{(1+h)^{-1}}^n_{n'}
(h^{l'}_{n\tilde{;}\beta}h^{n'\tilde{;}i'}_l-(2h^{l'}_{n\tilde{;}l}-
h^{l'}_{l\tilde{;}n})\cdot(h^{n'\tilde{;}i'}_{\beta}+
h^{n'i'}_{\tilde{;}\beta}-h^{i'\tilde{;}n'}_{\beta})-
2h^{l'}_{\beta\tilde{;}n}\cdot$$
$$\cdot(h^{n'i'}_{\tilde{;}l}-h^{i'\tilde{;}n'}_l))
-\frac1{4a^2}
(h^{i'}_{\beta,0}h^{l'}_{l,0}-2h^{i'}_{l,0}h^{l'}_{\beta,0})))
-\frac1{2a^2}h^{i'}_{\beta,0,0}-
\frac{a'}{a^3}h^{i'}_{\beta,0}+\frac{2\epsilon}{a^2}h^{i'}_\beta)-$$
$$-\frac{a'}{2a^3}\delta^\alpha_\beta{(1+h)^{-1}}^l_{l'}h^{l'}_{l,0},
\end{equation}
where $\epsilon=\pm1,0$  for the closed, open and
flat Universe. The sign "$\tilde{;}$" is used for
the covariant derivative in space part of the
metric and "$,0$" or comma for the derivative in
conformal time $\eta$. Metric $g^{ik}$ is defined
up to arbitrary coordinate transformations so one
can put the condition.

Solutions (\ref{raz}) for $h_{ik}$ can be written
as
$$h^i_k=S^i_k+V^i_k+B^i_k,$$ where
$S^i_k,V^i_k,B^i_k$ are irreducible scalar,
vector and tensor components of the tensor
satisfying the conditions \cite{Lif}:
$$\stackrel{(\circ)}g_{ik}S^{ik}=S\ne0,
\quad\stackrel{(\circ)}g_{ik}V^{ik}=0,
\quad\stackrel{(\circ)}g_{ik}B^{ik}=0.$$

Considering only the gravitational waves exclude
the scalar and vector parts by putting the gauge
conditions
\begin{equation}\label{sv}
h=0,\qquad{h^i_k}_{\tilde;i}=0.
\end{equation}

In this case the linearized equations for
perturbations $h^i_k$  take the form
\begin{equation}\label{lr}
h^{\alpha}_{\beta,0,0}+
\frac{2a'}ah^{\alpha}_{\beta,0}+\frac{2\epsilon}{a^2}h^\alpha_\beta+
h^{\alpha\tilde;\gamma}_{\beta\tilde;\gamma}=0.
\end{equation}

Go from the variables $h^i_k$ to new variables
$\mu^i_k=a(\eta)h^i_k$ and make the conformal
transformation $$\tilde g_{ik}=g_{ik}/a^2(\eta)$$

Then one obtains the equation for the field  with
spin 2 in Minkowsky flat space in some external
effective field
\begin{equation}\label{ure10}
\mu^\alpha_{\beta,0,0}+
\mu^{\alpha,\gamma}_{\beta,\gamma}-\frac{a''}{a}\mu^\alpha_\beta=0.
\end{equation}

After separation of variables and Fourier
representation $\mu^\alpha_\beta(x)$
\begin{equation}\label{ure11}
\mu^\alpha_\beta(x)=\int d^3k(g_{\vec
k}(\eta)e^{i\vec k\vec x}a^\alpha_\beta+g_{\vec
k}^*(\eta)e^{-i\vec k\vec x}a^{\alpha*}_\beta)
\end{equation}

one obtains for the time dependent $g_{\vec
k}(\eta)$ function the equation
\begin{equation}\label{ure12}
g_{\vec k}''(\eta)+(k^2-\frac{a''}a)g_{\vec
k}(\eta)=0,
\end{equation}
which formally has the negative square of the
effective mass $m^2a^2=-a''/a$.

\section{Third order equations for gravitational
waves} Let us consider the right hand side of
Einstein's equations. For (\ref{tei})
\begin{equation}\label{lr2}
\delta T^i_k=(\delta\varepsilon+\delta
p)u^iu_k-\delta^i_k\delta p+
(\varepsilon+p)(\delta u^iu_k+u^i\delta
u_k+\delta u^i\delta u_k).
\end{equation}

One has for the four velocities $u_i$ and
$\stackrel{(\circ)}u_i$ the conditions
$g_{ik}u^iu^k=1$ and
$\stackrel{(\circ)}g_{ik}\stackrel{(\circ)}u{}^i
\stackrel{(\circ)}u{}^k=1$. So
\begin{equation}\label{sk1}
2\stackrel{(\circ)}u{}_k\delta
u^k+\stackrel{(\circ)}g_{ik}\delta u^i\delta
u^k=0.
\end{equation}

Note that $\delta u_i$ due to constraints
(\ref{sv}) can depend only on
$h^j_kh^k_{j\tilde;i}$ \dots so $\delta
u^\alpha\delta u^\beta$ depends on the squares of
these terms. But we shall neglect the fourth and
higher orders. In synhronous reference system
$u^0=1/a,u^\alpha=0$ so from (\ref{sk1}) one has
$\delta u^0=0$ and
\begin{equation}\label{uh2}
\delta
R^0_0=\frac12\kappa(\delta\varepsilon+3\delta
p),\quad \delta
R^\alpha_\beta=\frac12\kappa\delta^\alpha_\beta(\delta
p-\delta\varepsilon),\quad \delta
R^\alpha_0=a\kappa(\varepsilon+p)\delta u^\alpha.
\end{equation}

Consider the case of flat space $\epsilon=0$.
Then $\displaystyle
h^\alpha_{\beta\tilde;\gamma}=
h^\alpha_{\beta,\gamma},h_\alpha^{\beta\tilde;\gamma}=
\frac1{a^2}h_\alpha^{\beta,\gamma}$ where Greek
indices are put up and below by use of the
Minkowsky metric. One can look on eqs.
(\ref{uh2}) as on Euler Lagrange equations for
the fields $h_{ik}$. Then due to constraints
(\ref{sv}) up to terms of the divergence form one
can obtain that not only $h^i_{k\tilde;i}=0$ but
$h^k_lh^i_{n\tilde;k}=0,\dots$ so that
(\ref{rue}) can be transformed to
\begin{equation}\label{ruesv}
\delta R^\alpha_\beta={(1+h)^{-1}}^\alpha_{i'}
(\frac12{(1+h)^{-1}}^l_{l'} (
-h^{i'\tilde{;}l'}_{\beta\tilde{;}l}-
h^{l'\tilde{;}i'}_{l\tilde{;}\beta})+$$
$$+\frac14{(1+h)^{-1}}^l_{l'} {(1+h)^{-1}}^n_{n'}
(h^{l'}_{n\tilde{;}\beta}h^{n'\tilde{;}i'}_l-h^{l'}_{l\tilde{;}n}
h^{i'\tilde{;}n'}_{\beta}+
2h^{l'}_{\beta\tilde{;}n}h^{i'\tilde{;}n'}_l))-$$
$$-\frac{a'}{2a^3}\delta^\alpha_\beta{(1+h)^{-1}}^l_{l'}h^{l'}_{l,0}+
{(1+h)^{-1}}^{\alpha}_{i'}(-\frac1{2a^2}h^{i'}_{\beta,0,0}-
\frac{a'}ah^{i'}_{\beta,0}-$$
$$-\frac1{4a^2}{(1+h)^{-1}}^l_{l'}
(h^{i'}_{\beta,0}h^{l'}_{l,0}-2h^{i'}_{l,0}h^{l'}_{\beta,0})).
\end{equation}

From (\ref{ruesv}) follows that one can take
instead of (\ref{uh}) the equations
\begin{equation}\label{url}
(1+h)^\gamma_\alpha\delta
R^\alpha_\beta=\frac12\kappa(1+h)^\gamma_\beta(\delta
p-\delta\varepsilon).
\end{equation}

Multiply the last equation on "$-2a^2$", then
from (\ref{ruesv}), (\ref{url}) one obtains
\begin{equation}\label{uregl}
h^\alpha_{\beta,0,0}+\frac{2a'}ah^\alpha_{\beta,0}
+\frac{a'}{a}(1+h)^\alpha_\beta{(1+h)^{-1}}^l_{l'}
h^{l'}_{l,0}+$$
$$+\frac12{(1+h)^{-1}}^l_{l'}
(h^\alpha_{\beta,0}h^{l'}_{l,0}-
2h^\alpha_{l,0}h^{l'}_{\beta,0})
+{(1+h)^{-1}}^l_{l'} (h^{\alpha;l'}_{\beta;l}+
h^{l',\alpha}_{l,\beta})-$$
$$-\frac12{(1+h)^{-1}}^l_{l'} {(1+h)^{-1}}^n_{n'}
(h^{l'}_{n,\beta}h^{n',\alpha}_l- h^{l'}_{l,n}
h^{\alpha,n'}_{\beta}+
2h^{l'}_{\beta,n}h^{\alpha,n'}_l)=$$
$$=a^2\kappa(1+h)^\alpha_\beta(\delta\varepsilon-\delta
p).
\end{equation}

Consider first three orders in  $h^i_k$  in
equations (\ref{uregl}).
\begin{equation}\label{uregl2}
h^\alpha_{\beta,0,0}+\frac{2a'}ah^\alpha_{\beta,0}+
\frac12(\delta^l_{l'}-h^l_{l'}+ h^l_kh^k_{l'})
(h^\alpha_{\beta,0}h^{l'}_{l,0}-
2h^\alpha_{l,0}h^{l'}_{\beta,0}
+2h^{\alpha,l'}_{\beta,l}+
2h^{l',\alpha}_{l,\beta}-$$
$$-(\delta^n_{n'}-h^n_{n'})
(h^{l'}_{n,\beta}h^{n',\alpha}_l- h^{l'}_{l,n}
h^{\alpha,n'}_{\beta}+
2h^{l'}_{\beta,n}h^{\alpha,n'}_l))=$$
$$=(\delta^\alpha_\beta+h^\alpha_\beta)(a^2\kappa(\delta\varepsilon-\delta
p)+\frac{a'}{a}(h^l_{l'}-h^l_kh_{l'}^k)h^{l'}_{l,0}),
\end{equation}
or
\begin{equation}\label{uregl3}
h^\alpha_{\beta,0,0}+\frac{2a'}ah^\alpha_{\beta,0}+
h^{\alpha,l}_{\beta,l}-h^\alpha_{l,0}h^{l}_{\beta,0}
-h^\alpha_{l,n}h^{l,n}_\beta-\frac12h^{n,\alpha}_lh^l_{n,\beta}
-h^l_{l'}h^{l',\alpha}_{l,\beta}-$$
$$-\delta^\alpha_\beta(a^2\kappa(\delta\epsilon-\delta p)+\frac{a'}a
h^n_{n'}h^{n'}_{n,0})+$$
$$+\frac12h^l_{l'}(
(2h^{l'}_{n,\beta}h^{n,\alpha}_l-2h^{\alpha,n}_lh^{l'}_{\beta,n}
-h^{l'}_{l,n}h^{\alpha,n}_\beta+2h^{l'}_lh^{l,\alpha}_{l',\beta}
-h^\alpha_{\beta,0}h^{l'}_{l,0}+2h^{l'}_{\beta,0}h^\alpha_{l,0}+
$$
$$+\frac{2a'}{a}\delta^\alpha_\beta h_l^nh^{l'}_{n,0}-
\frac{2a'}{a}h^\alpha_\beta)h^{l'}_{l,0})=
h^\alpha_\beta a^2\kappa(\delta\varepsilon-\delta
p)=0.
\end{equation}

So
\begin{equation}\label{uregl4}
\delta^\alpha_\beta a^2\kappa(\delta
p-\delta\varepsilon)=
h^\alpha_{l,0}h^{l}_{\beta,0}
+h^\alpha_{l,n}h^{l,n}_\beta+\frac12h^{n,\alpha}_lh^l_{n,\beta}
+h^l_{l'}h^{l',\alpha}_{l,\beta}+\delta^\alpha_\beta\frac{a'}a
h^n_{n'}h^{n'}_{n,0}
\end{equation}

Putting away the divergence of
$h^\alpha_{l,\gamma}h^{l,\gamma}_\beta+
\frac12h^{\gamma,\alpha}_lh^l_{\gamma,\beta}
+h^l_{l'}h^{l',\alpha}_{l,\beta}$ and taking into
account for fixed nonzero components of the
tensor $h^i_k$ the condition
$h^l_{l'}h^n_lh^{l'}_n=0$ after simple
transformations one obtains
\begin{equation}\label{uregl5}
h^\alpha_{\beta,0,0}+\frac{2a'}ah^\alpha_{\beta,0}+
h^{\alpha,\gamma}_{\beta,\gamma}+h^l_{l'}h^{l'}_{\beta,0}h^{\alpha}_{l,0}+$$
$$+h^l_{l'}(
\frac12h^{l'}_{l,n}h^{\alpha,n}_\beta+h^{l',\alpha}_n
h^n_{l,\beta}-h^{l'}_{\beta,n}h^{\alpha,n}_l+
h^l_nh_{l',\beta}^{n,\alpha})=0.
\end{equation}

Now let us go from variables $h^i_k$ to variables
$\mu^i_k=a(\eta)h^i_k$ and make the conformal
transformation $$\tilde
g_{ik}=g_{ik}/a^2(\eta).$$

Then we obtain the equation in flat Minkowsky
space with some effective external field
\begin{equation}\label{ure3}
\mu^\alpha_{\beta,0,0}+
\mu^{\alpha,\gamma}_{\beta,\gamma}-
\frac{a''}{a}\mu^\alpha_\beta+
\frac1{a^2}(\mu^\alpha_{l,0}-\frac{a'}a\mu^\alpha_l)
(\mu^n_{\beta,0}-\frac{a'}a\mu^n_\beta)\mu^l_n+$$
$$+\frac1{a^2}\mu^l_{l'}(
\frac12\mu^{l'}_{l,n}\mu^{\alpha,n}_\beta+\mu^{l',\alpha}_n
\mu^n_{l,\beta}-\mu^{l'}_{\beta,n}\mu^{\alpha,n}_l+
\mu^l_n\mu_{l',\beta}^{n,\alpha})=0.
\end{equation}

\section{Spontaneous breaking of symmetry for gravitons}
Let us consider vacuum solution (\ref{ure3})
depending only on time. In quantum field theory
this means dependence of vacuum on time. Then
\begin{equation}\label{uret11}
\mu^\alpha_{\beta,0,0}=\frac{a''}{a}\mu^\alpha_\beta-
\frac{a'^2}{a^4}\mu^\alpha_l\mu^l_n\mu^n_\beta.
\end{equation}

Taking into account constraints (\ref{sv}) in
variables $\mu^1_1=-\mu^2_2,
\quad\mu^1_2=\mu^2_1$ one gets the potential
corresponding to (\ref{uret11}) as

\begin{equation}\label{cal3}
V=-\frac{a''}{2a}\mu^\alpha_\beta\mu_\alpha^\beta+\frac{a'^2}{8a^4}
(\mu^\alpha_\beta\mu_\alpha^\beta)^2.
\end{equation}

Write the field $\mu^\alpha_\beta$ close to the
minimum of the potential energy
\begin{equation}\label{cal4}
\frac{a''}a=\frac{a'^2}{2a^4}\mu^\alpha_\beta(0)\mu_\alpha^\beta(0),\quad
\mu^\alpha_\beta=\mu^\alpha_\beta(0)+\xi^\alpha_\beta.
\end{equation}

One must note that the condition (\ref{cal4}) on
$\mu^\alpha_\beta(0)$ is the condition of minimal
energy at same fixed moment $t_0$.

This is basic idia. Initied of dealing this time
depended $m,\lambda$ we put the initial
conditions at some $t_0$. Dusing these
coordinations form the principle minimal energy
at this moment. Surely $m,\lambda$ at this moment
are numbers.

Take the solution (\ref{cal4}) as
\begin{equation}\label{calm}
\mu^1_2(0)=\mu^2_1(0)=0,\quad
\mu^1_1(0)=-\mu^2_2(0)=\mu_0=\sqrt{\frac{a''a^3}{a'^2}}.
\end{equation}

Then the Lagrangian
$$L=\frac12\mu^{\alpha,n}_\beta\mu_{\alpha,n}^\beta+
\frac{a''}{2a}\mu^\alpha_\beta\mu_\alpha^\beta-\frac{a'^2}{8a^4}
(\mu^\alpha_\beta\mu_\alpha^\beta)^2$$ can be written as
$$L=\frac12(\mu^\alpha_\beta(0)+\xi^\alpha_\beta)^{,n}
(\mu_\alpha^\beta(0)+\xi_\alpha^\beta)_{,n}+
\frac{a''}{2a}(\mu^\alpha_\beta(0)+\xi^\alpha_\beta)
(\mu_\alpha^\beta(0)+\xi_\alpha^\beta)-$$
$$-\frac{a'^2}{8a^4}(\mu^\alpha_\beta(0)\mu_\alpha^\beta(0)
+\xi^\alpha_\beta\xi_\alpha^\beta+
2\mu_\alpha^\beta(0)\xi_\alpha^\beta)^2=$$
$$=\mu_{0,0}^2+\frac12\xi^{\alpha,n}_l\xi_{\alpha,n}^l+
\mu_{0,0}(\xi^1_{1,0}-\xi^2_{2,0})+
\frac{a''}{2a}(2\mu_0^2+2\mu_0(\xi^1_1-
\xi^2_2)+\xi^\alpha_\beta\xi_\alpha^\beta)-$$
$$-\frac{a'^2}{8a^4}(4\mu_0^4+(\xi^\alpha_\beta\xi_\alpha^\beta)^2
+4\mu_0\xi^\alpha_\beta\xi_\alpha^\beta(\xi^1_1-\xi^2_2)+8\mu_0^3(\xi_1^1-\xi_2^2)+$$
$$+4\mu_0^2\xi^\alpha_\beta\xi_\alpha^\beta+
4\mu_0^2((\xi_1^1)^2+(\xi_2^2)^2-2\xi_1^1\xi^2_2)).$$

Consider the quadratic in $\xi^\alpha_\beta$
terms
\begin{equation}\label{lag1}
L=\frac12\xi^{\alpha,n}_l\xi_{\alpha,n}^l+
\frac{a''}{2a}\xi^\alpha_\beta\xi_\alpha^\beta-
\frac{a'^2}{2a^4}\mu_0^2(\xi_\beta^\alpha\xi^\beta_\alpha+
(\xi_1^1)^2+(\xi_2^2)^2-2\xi_1^1\xi^2_2).
\end{equation}

Taking into account the equality
$\displaystyle\mu_0^2\frac{a'^2}{a^4}=\frac{a''}a$ and the gauge
\hbox{$\xi^1_1+\xi^2_2=0$} one obtains the Lagrangian for gravitons
\begin{equation}\label{lag2}
L(\xi)=\frac12\xi^{\alpha,n}_\beta\xi_{\alpha,n}^\beta-
\frac{a''}a((\xi_1^1)^2+(\xi^2_2)^2),
\end{equation}
which is called by us the effective Lagrangian
after the spontaneous breaking of symmetry.
                                                                                                               (46).
Euler-Lagrange equations have the form
\begin{equation}\label{lag3}
\xi^{1,n}_{1,n}+\frac{2a''}a\xi^1_1=0,\quad
\xi^{2,n}_{2,n}+\frac{2a''}a\xi^2_2=0,\quad
\xi^{2,n}_{1,n}=0,\quad \xi^{1,n}_{2,n}=0.
\end{equation}

One sees that now in (\ref{lag3}) the sign  of
the mass squared is a correct one. The components
$\xi^1_2,\xi^2_1$ are massless while
$\xi^1_1,\xi^2_2$ have the nonnegative mass
squared $\displaystyle\frac{2a''}a$. The
solutions for diagonal components
$\xi^1_1,\xi^2_2$ (or
$\xi^\alpha_\alpha,\alpha=1,2$) can be written as
$\xi$ the Fourier integral
\begin{equation}\label{Fur1}
\xi(x)=\frac1{(2\pi)^3}\int d\vec k(c_{\vec
k}g_k^*(\eta)e^{i\vec k\vec x}+c^*_{\vec k}
g_k(\eta)e^{-i\vec k\vec x}),
\end{equation}

And one has the equation
\begin{equation}\label{vr1}
g''_k+(k^2+\frac{2a''}a)g_k=0.
\end{equation}

This equation is free from the problem of the negative square of the
effective mass if $\displaystyle\frac{2a''}a>0$ and one can
construct the quantum theory of gravitons based on the new vacuum
state (\ref{calm}). One can notice that the vacuum expectation value
of the field $\mu(\eta)$  is close to the scale factor. For the
scale factor $a(\eta)=\eta^p$ one obtains
$$\mu_0=a(\eta)\sqrt{\frac{p-1}p}.$$

One sees that for all $p\in[-1;0)$ (if
$p\in(0;1)$ then $m^2=-2a''/a>0$ and we leave
vacuum as $\mu_0=0$) the dynamical perturbation
$h^\alpha_\beta\geqslant1$ which contradicts the
condition of the expansion of the curvature
tensor into a series in this perturbation. The
value $p=-1$ corresponds to inflation, so we
cannot deal the inflation model here while other
situations can be considered. However this case
must be considered separately and it is not
studied in this paper.

\section{The Lagrange formalism for gravitons}
One can see from (\ref{Fur1}-\ref{vr1}) that gravitons in the
expanding isotropic Universe are described by the effective scalar
field $\vartheta(x)$ with the Lagrangian
\begin{equation}\label{lk1}
L=\sqrt{-g}(\vartheta(x)_{,n}\vartheta(x)^{,n}-\frac13R\vartheta(x)\vartheta(x)),
\end{equation}
where $g=\det(g_{ik})$ and $R$ the scalar
curvature. There is no term $\frac12$ because one
deals with two polarizations. Euler-Lagrange
equation for the field is
$\xi^1_1=\xi^2_2=\xi=\vartheta\cdot a$. Look for
solutions of (\ref{vr1}) in the form
(\ref{Fur1}). The numbers $c_{\vec k},c^*_{\vec
k}$ are changed on the operators $\widehat
c_{\vec k},\widehat c{\,}^+_{\vec k}$ with
commutation relations
\begin{equation}\label{kom1}
[\widehat c_{\vec k},\widehat c{\,}^+_{\vec
k'}]=\delta(\vec k-\vec k'),\quad [\widehat
c_{\vec k},\widehat c_{\vec k'}]=[\widehat
c{\,}^+_{\vec k},\widehat c{\,}^+_{\vec k'}]=0.
\end{equation}

The Fock vacuum state $\rm|in>$ is defined as
$$\widehat c_{\vec k}\rm|in>=0,\quad<in|in>=1.$$

Then for $\displaystyle\widehat\vartheta(x)=\frac1a\widehat\xi(x)$
one has
\begin{equation}\label{Fur2}
\widehat\vartheta(x)=\frac1{(2\pi)^3a(\eta)}\int d\vec k(\widehat
c_{\vec k}g_k^*(\eta)e^{i\vec k\vec x}+\widehat c{\,}^+_{\vec
k}g_k(\eta)e^{-i\vec k\vec x}),
\end{equation}
where $g_k(\eta)$  satisfy (\ref{vr1}) written as
\begin{equation}\label{vr2}
g''_k+\omega_k^2(\eta)g_k=0,\quad\omega_k^2(\eta)=\frac{2a''}a+k^2
\end{equation}
with initial conditions
\begin{equation}\label{vr3}
g_k(\eta_0)=\frac1{\sqrt{\omega_k(\eta_0)}},\quad
g'_k(\eta_0)=i\sqrt{\omega_k(\eta_0)}.
\end{equation}

The condition for the  wronskian
\begin{equation}\label{vr4}
g_k(\eta_0){g'_k}^*(\eta_0)-g_k^*(\eta_0)g'_k(\eta_0)=-2i
\end{equation}
leads to existence of the full set of solutions
of (\ref{vr1}) in the sense of the indefinite
scalar product
\begin{equation}\label{vr5}
(\xi_1,\xi_2)=i\int d\vec
x(\xi_1^*\stackrel{\longleftrightarrow}{\partial_0}\xi_2).
\end{equation}

The Hamiltonian of the quantized field
$\widehat\xi(x)$ in the metric (\ref{met}) has
the form
\begin{equation}\label{vr6}
\widehat H(\eta)=\int d\vec
x(\widehat\xi'^+\widehat\xi'+\frac{2a''}a\widehat\xi^+\widehat\xi).
\end{equation}

Putting the field $\widehat\vartheta(x)$ from (\ref{vr6}) into
(\ref{Fur2}) one obtains
$$\widehat H(\eta)=\frac1{\pi^2a^4(\eta)}\int_0^\infty k^2dk\omega_k(\eta)
(E_k(\eta)(\widehat c{\,}^+_k\widehat c_k+
\widehat c{\,}_k\widehat c_k^+)+$$
\begin{equation}\label{vr7}
+F_k\widehat c{\,}^+_k\widehat c{\,}^+_k
+F_k^*\widehat c{\,}_k\widehat c{\,}_k),
\end{equation}

where the coefficients $E_k,F_k$ are expressed
through the solutions of the (\ref{vr2})
\begin{equation}\label{vr7}
E_k(\eta)=\frac1{2\omega}(|g_k'|^2+|g_k|^2)\qquad
F_k(\eta)=\frac1{2\omega}(g_k'^2+g_k^2).
\end{equation}

The corpuscular interpretation can be made in
terms of creation and annihilation operators
$\widehat b{\,}_k,\widehat b{\,}^+_k$
diagonalizing the Hamiltonian. If
$$\widehat c{\,}_k=\alpha_k^*(\eta)\widehat b{\,}_k -\beta_k(\eta)\widehat
b{\,}^+_k$$

then the Hamiltonian is
\begin{equation}\label{dgm}
\widehat
H(\eta)=\frac1{\pi^2a^4(\eta)}\int_0^\infty
k^2dk\omega_k(\eta) (E_k(\eta)-1)(\widehat
b{\,}^+_k\widehat b_k+ \widehat b{\,}_k\widehat
b_k^+).
\end{equation}

The density of created particles and their energy
density \cite{Grib} can be found using formulas
\begin{equation}\label{plch}
n(\eta)=\frac1{\pi^2a^3(\eta)}\int_0^\infty
k^2dk|\beta_k|^2.
\end{equation}

\section{Some models of  graviton creation}
Let us consider some matter filling the Universe
with the equation of state $p=\gamma\varepsilon$
where $p$ is pressure and $\varepsilon$ the
energy density. One has for the homogeneous
quasieuclidean isotropic Universe the equation
\cite{Lan}
\begin{equation}\label{ed}
\frac{8\pi\kappa}{c^4}\varepsilon=\frac{3a'^2}{a^4},
\quad\hbox{than}\quad
a(\eta)=C\eta^{\frac2{1+3\gamma}},
\end{equation}

Let us take $a(\eta)=C\eta^p$ and put it into
(\ref{vr2}), then one obtains (However one must
remember that due to our condition not all p can
be used. One must have p>1.)
\begin{equation}\label{mst1}
g''(\eta)+(2p(p-1)\frac1{\eta^2}+k^2)g(\eta)=0.
\end{equation}
where
$$m^2=2p(p-1)=\frac{4(1-3\gamma)}{(1+3\gamma)^2},
\quad a(\eta)=\frac1\eta.$$

So the results obtained for the scalar field in
\cite{Grib} are  valid for gravitons for any
scale factor with the scale factor of a given
form. In \cite{Grib} it was shown that for the
density of created particles and the energy
density defined by (\ref{dgm} -- \ref{plch}) one
gets convergent integrals. Let us calculate them.
Putting the notation $x=k\eta$ one gets
\begin{equation}\label{mst2}
\frac{d^2g}{dx^2}+(1+\frac{m^2}{x^2})g(x)=0.
\end{equation}

Then the energy density of created particles due
to (\ref{mst2}) with is calculated as
$$\varepsilon(\eta)=<0|T_0^0|0>=$$
\begin{equation}\label{plen1}
=\frac2{\pi^2(a(\eta)\eta)^4}\int_0^\infty x^3dx
\omega(x)(\frac1{2\omega(x)}(|\frac{dg(x)}{dx}|^2+\omega^2(x)|g(x)|^2)-1)
\end{equation}

The solutions of (\ref{mst2}) are Bessel
functions
$$g(x)=C_1\sqrt{\frac{\pi x}2}J(\frac12\sqrt{1-4m^2},x)+
C_2\sqrt{\frac{\pi
x}2}Y(\frac12\sqrt{1-4m^2},x).$$

Than
\begin{equation}\label{plen2}
\varepsilon(\eta)\thickapprox\frac2{\pi^2(a(\eta)\eta)^4}0.04m^3=
\frac{1.5\cdot10^{-3}R^{3/2}}{a(\eta)\eta^4},\qquad
0<m^2<4.
\end{equation}

For the density of created particles in the unit
volume one gets
\begin{equation}\label{plchm2}
n(\eta)=\frac1{\pi^2(a(\eta)\eta)^3}\int_0^\infty
x^2dx
(\frac1{2\omega(x)}(|\frac{dg(x)}{dx}|^2+\omega^2(x)|g(x)|^2)-1),
\end{equation}

This integral is convergent \cite{Grib}. For
small $m$ ($0<m<0.5$) $n(\eta)\sim R$ and for
large $m$ ($m>0.5$) $n(\eta)\sim\sqrt R.$\par
Consider dust Universe with $a(\eta)=C\eta^2$.
Then

\begin{equation}\label{et1}
\varepsilon_{grav.}=\frac{4\cdot10^{-3}}{t^4}\quad(c^{-4})
\end{equation}

For the background classical matter one has
\begin{equation}\label{et2}
\varepsilon_{matt.}=\frac{2\cdot10^{84}cek^{-2}}{t^2}\quad(c^{-4})
\end{equation}

So for Planckean time ($t_{pl}=10^{-43}cek$) the
graviton energy density created from vacuum is
some ten percent of the matter density while at
the inflation time $t_{inf}=10^{-36}cek$) it is
only $10^{-14}$ of matter. These numbers are
consistent with our approximation for the metric
in perturbation theory. At the modern epoch one
gets from (\ref{et1}) that the energy flow from
the time of the end of inflation
$t_{inf}=10^{-36}cek$ is
\begin{equation}\label{et3}
\varepsilon_{sov. grav.}=0.5\cdot10^{-12}\quad(\frac{erg}{s\cdot
cm^2})
\end{equation}

This can be compared with the flow from the Crab nebula \cite{Vein}.
One sees that it is much smaller.
\begin{equation}\label{et4}
\varepsilon_{krab}=10^{-8}\quad(\frac{erg}{s\cdot cm^2})
\end{equation}

\section{Acknowledgements}
The authors are indebted to the participants of the A. A. Friedmann
seminar of St. Petersburg for the discussions of the paper and to
Ministry of Education and Science of Russia (grant RNP.2.1.6826) for
financial support.

\end{document}